\documentclass[preprint,12pt]{elsarticle}

\usepackage{graphics}
\usepackage{graphicx}
\usepackage{epsfig}

\usepackage{amssymb}
\usepackage{amsthm}

\usepackage{amsmath}
\usepackage{supertabular} 
\usepackage{placeins}

\def\tstrut{\vrule height2.5ex depth0pt width0pt} 

\biboptions{square,sort&compress}

\journal{Nuclear Physics A}

\begin{document}

\begin{frontmatter}



\title{Strong charmonium decays in a microscopic model}


\author[SAL]{J. Segovia\corref{ARG}}
\author[SAL]{D.R. Entem}
\author[SAL]{F. Fern\'andez}

\address[SAL]{Grupo de F\'{\i}sica Nuclear and IUFFyM, Universidad de Salamanca,
E-37008 Salamanca, Spain}

\cortext[ARG]{Current address: Physics Division, Argonne
National Laboratory, Argonne, IL 60439, USA.}

\begin{abstract}
Although the spectra of heavy quarkonium systems have been successfully explained by
certain QCD motivated potential models, their strong decays are still an open
problem. We perform a microscopic calculation of vector charmonium strong decays into
open-charm mesons where the $q\bar{q}$ pairs are created from the same interquark
interactions acting in the quark model that has been used to describe its spectrum,
and also its leptonic and radiative decays. We compare the numerical results with
those predicted by the $^{3}P_{0}$ decay model and with the available experimental
data, and discuss the possible influence on the strong widths of the different terms
of the potential. A comparison with other predictions from similar microscopic decay
models is also included.
\end{abstract}


\end{frontmatter}


\section{Introduction}
\label{sec:introduction}

Since its discovery in 1974~\cite{aub74_1,aug74_1}, the charmonium system has become
the prototypical 'hydrogen atom' of meson spectroscopy~\cite{app75_1,eic76_1}. The
spectrum of relatively narrow states below the open-charm threshold at $3.73\,{\rm
GeV}$ can be identified with the $1S$, $1P$ and $2S$ $c\bar{c}$ levels predicted by
potential models which incorporate a color Coulomb term at short distances and a
color confining term at large distances. Difficulties arises above the open-charm 
threshold due to the coupling with the meson-meson continuum.

In recent times the construction of $B$-factories has triggered the discovery of
many new particles, denoted as $XYZ$ mesons, whose nature supposes a challenge for
the theorists and can contribute to a better knowledge of the $c\bar{c}$
phenomenology.

One open topic of great interest on charmonium states is their strong decays which
constitute a rather poorly understood area of hadronic physics. A great part of our
knowledge of strong interaction comes from decays and for that reason it is important
to pursue the most complete description of them.

Attempts of modeling strong decays date from Micu's suggestion~\cite{mic69_1} that
hadron decays proceed through $q\bar{q}$ pair production with vacuum quantum numbers,
$J^{PC}=0^{++}$. Since this corresponds to a $^{3}P_{0}$ $q\bar{q}$ state, it is now
generally referred to as the $^{3}P_{0}$ decay model. This suggestion was developed
and applied extensively by Le Yaouanc {\it et al.}~\cite{yao73_1-1,yao73_1-2} in the
1970s. Studies of hadron decays using the $^{3}P_{0}$ model have been concerned
almost exclusively with numerical predictions, and have not led to any fundamental
modifications to the original model. Recent studies have considered changes in the
spatial dependence of the pair production amplitude as a function of quark
coordinates~\cite{tot96_1,gei94_1,kok87_1,god85_1}. There have been some studies
of the decay mechanism which consider an alternative phenomenological model in which
the $q\bar{q}$ pair is produced with $^{3}S_{1}$ quantum numbers~\cite{alc84_1}.
However, this possibility seems to disagree with experiment~\cite{gei94_1}.

An alternative procedure is the study of strong decays through microscopic decay
models. The difference between this approach and those described above lies on the
description of the $q\bar{q}$ pair creation vertex. In the microscopic decay models
the $q\bar{q}$ pair is created from interquark interactions acting in the quark
model. The differences between calculations of this kind lies in the choice of the
pieces of the potential which enter in the vertex calculation.

So Eichten {\it et al.}~\cite{Eichten78,Eichten06} assumes that the $q\bar{q}$
production is due to the time-like part of the vector Lorentz confining interaction,
while Ackleh {\it et al.}~\cite{Swanson96} and Bao {\it et al.} ~\cite{Bao-Fei2011}
assumes that the $q\bar{q}$ pair is produced by one-gluon exchange and scalar
confining interactions. Then, the description of the strong decays is intimately
related with the problem of the Dirac structure of the confinement. 

From the point of view of the hadronic spectroscopy, confinement has to be dominantly
scalar in order to reproduce the hyperfine splittings observed in heavy
quarkonium~\cite{Dobbs2008,Lees2011,Adachi2011}, although to explain the light quark 
phenomenology a small mixture (of the order of $20\%$) of vector confinement is
needed ~\cite{Vijande2005}. However, build Hamiltonian-based models of
QCD~\cite{PhysRevD.55.1578} seems to require vector confinement.

In the present work, we generalize the microscopic decay models of references
mentioned above using the  quark interaction of Ref.~\cite{Vijande2005} which
includes one gluon exchange plus a mixture of scalar and vector confinement. This
model successfully describes hadron phenomenology and hadronic reactions and has
recently been applied to mesons containing heavy quarks in
Refs.~\cite{Segovia2008,Segovia2009}. We study the open-charm strong decays of the
$1^{--}$ $c\bar c$ resonances looking for the possible influence of the mixture
of scalar and vector Lorentz structure.

The content of the present paper is organized as follows. In
section~\ref{sec:quarkmodel} we briefly explain the main features of the 
constituent quark model. Section~\ref{sec:decaymodel} is devoted to the microscopic
description of the strong decay mechanism. Section~\ref{sec:comparison} is dedicated
to show our results and how its compared with those of the $^{3}P_{0}$ model. A
comparison of the numerical values coming from our microscopic decay model with those
of other similar microscopic decay models is also included. We finish in
Section~\ref{sec:conclusions} with some remarks and conclusions.

\section{Constituent quark model}
\label{sec:quarkmodel}

One consequence of the spontaneous chiral symmetry breaking is that the nearly
massless ``current'' light quarks acquire a dynamical, momentum-dependent mass $M(p)$
with $M(0)\approx 300\,{\rm MeV}$ for the $u$ and $d$ quarks, namely, the constituent
mass. To preserve chiral invariance of the QCD Lagrangian new interaction terms,
given by Goldstone-boson exchanges, should appear between constituent quarks. This
together with the perturbative one-gluon exchange (OGE) and the nonperturbative
confining interactions are the main pieces of our potential model~\cite{Vijande2005}.

The wide energy range covered by a consistent description of light, strange and heavy
mesons requires an effective scale-dependent strong coupling constant. We use the
frozen coupling constant of Ref.~\cite{Vijande2005}
\begin{equation}
\alpha_{s}(\mu)=\frac{\alpha_{0}}{\ln\left(
\frac{\mu^{2}+\mu_{0}^{2}}{\Lambda_{0}^{2} } \right)},
\end{equation}
where $\mu$ is the reduced mass of the $q\bar{q}$ pair and $\alpha_{0}$, $\mu_{0}$
and $\Lambda_{0}$ are parameters of the model determined by a global fit to all meson
spectrum.

In the heavy quark sector chiral symmetry is explicitly broken and Goldstone-boson
exchanges do not appear. Only the OGE and confinement potentials are present and
contain central, tensor and spin-orbit contributions. For the OGE potential, they are
given by
\begin{equation}
\begin{split}
&
V_{\rm OGE}^{\rm C}(\vec{r}_{ij}) =
\frac{1}{4}\alpha_{s}(\vec{\lambda}_{i}^{c}\cdot
\vec{\lambda}_{j}^{c})\left[ \frac{1}{r_{ij}}-\frac{1}{6m_{i}m_{j}} 
(\vec{\sigma}_{i}\cdot\vec{\sigma}_{j}) 
\frac{e^{-r_{ij}/r_{0}(\mu)}}{r_{ij}r_{0}^{2}(\mu)}\right], \\
& 
V_{\rm OGE}^{\rm T}(\vec{r}_{ij})=-\frac{1}{16}\frac{\alpha_{s}}{m_{i}m_{j}}
(\vec{\lambda}_{i}^{c}\cdot\vec{\lambda}_{j}^{c})\left[ 
\frac{1}{r_{ij}^{3}}-\frac{e^{-r_{ij}/r_{g}(\mu)}}{r_{ij}}\left( 
\frac{1}{r_{ij}^{2}}+\frac{1}{3r_{g}^{2}(\mu)}+\frac{1}{r_{ij}r_{g}(\mu)}\right)
\right]S_{ij}, \\
&
\begin{split}
V_{\rm OGE}^{\rm SO}(\vec{r}_{ij})= &  
-\frac{1}{16}\frac{\alpha_{s}}{m_{i}^{2}m_{j}^{2}}(\vec{\lambda}_{i}^{c} \cdot
\vec{\lambda}_{j}^{c})\left[\frac{1}{r_{ij}^{3}}-\frac{e^{-r_{ij}/r_{g}(\mu)}}
{r_{ij}^{3}} \left(1+\frac{r_{ij}}{r_{g}(\mu)}\right)\right] \times \\ & \times 
\left[((m_{i}+m_{j})^{2}+2m_{i}m_{j})(\vec{S}_{+}\cdot\vec{L})+
(m_{j}^{2}-m_{i}^{2}) (\vec{S}_{-}\cdot\vec{L}) \right],
\end{split}
\end{split}
\end{equation}
where $\vec{S}_{\pm}=\frac{1}{2}(\vec{\sigma}_{i}\,\pm\,\vec{\sigma}_{j})$. Besides,
$r_{0}(\mu)=\hat{r}_{0}\frac{\mu_{nn}}{\mu_{ij}}$ and
$r_{g}(\mu)=\hat{r}_{g}\frac{\mu_{nn}}{\mu_{ij}}$ are regulators which depend on
$\mu_{ij}$, the reduced mass of the interacting quarks. The contact term of the
central potential of one-gluon exchange has been regularized in a suitable way as
\begin{equation}
\delta(\vec{r}_{ij})\sim\frac{1}{4\pi
r_{0}^{2}}\frac{e^{-r_{ij}/r_{0}}}{r_{ij}}.
\end{equation}

The breaking of the color electric string between two static color sources is a
phenomenon predicted by QCD and it is the basis of the meson decays and hadronization
processes. Although there is no analytical proof, it is a general belief that
confinement emerges from the force between the gluon color charges. When two quarks
are separated, due to the non-Abelian character of the theory, the gluon fields
self-interact forming color strings which bring the quarks together.

In a pure gluon gauge theory the potential energy of the $q\bar{q}$ pair grows
linearly with the quark-antiquark distance. However, in full QCD the presence of
sea quarks soften the linear potential, due to the screening of the color
charges, and eventually leads to the breaking of the string. We incorporate it
in our confinement potential as
\begin{equation}
\begin{split}
&
V_{\rm CON}^{\rm C}(\vec{r}_{ij})=\left[-a_{c}(1-e^{-\mu_{c}r_{ij}})+\Delta
\right] (\vec{\lambda}_{i}^{c}\cdot\vec{\lambda}_{j}^{c}), \\
&
\begin{split}
V_{\rm CON}^{\rm SO}(\vec{r}_{ij})= &
-(\vec{\lambda}_{i}^{c}\cdot\vec{\lambda}_{j}^{c}) \frac{a_{c}\mu_{c}e^{-\mu_{c}
r_{ij}}}{4m_{i}^{2}m_{j}^{2}r_{ij}}\left[((m_{i}^{2}+m_{j}^{2})(1-2a_{s})
\right. \\ & \left.
+4m_{i}m_{j}(1-a_{s}))(\vec{S}_{+}\cdot\vec{L})+(m_{j}^{2}-m_{i}^{2}) (1-2a_{s})
(\vec{S}_{-}\cdot\vec{L}) \right],
\end{split}
\end{split}
\end{equation}
where $a_{s}$ controls the ratio between the scalar and vector Lorentz structure
\begin{equation}
V_{\rm CON}(\vec{r}_{ij})=a_{s}V_{\rm
CON}^{\rm scalar}(\vec{r}_{ij})+(1-a_{s})V_{\rm CON}^{\rm
vector}(\vec{r}_{ij}).
\end{equation}

At short distances this potential presents a linear behavior with an effective
confinement strength $\sigma=-a_{c}\,\mu_{c}\,(\vec{\lambda}^{c}_{i}\cdot 
\vec{\lambda}^{c}_{j})$ while it becomes constant at large distances. This 
type of potential shows a threshold defined by
\begin{equation}
V_{thr}=\{-a_{c}+ \Delta\}(\vec{\lambda}^{c}_{i}\cdot \vec{\lambda}^{c}_{j}).
\end{equation}

No $q\bar{q}$ bound states can be found for energies higher than this threshold and
the system suffers a transition from a color string configuration between two static
color sources into a pair of static mesons due to the breaking of the color flux-tube
and the most favored subsequent decay into hadrons.

To find the quark-antiquark bound states with these interactions, we solve the
Schr\"odinger equation using the Gaussian Expansion Method~\cite{Hiyama2003}. It 
allows us to evaluate easily the strong decay amplitudes.

In this method the radial wave function, solution of the Schr\"odinger equation, is
expanded in terms of basis functions as
\begin{equation}
R_{\alpha}(r)=\sum_{n=1}^{n_{max}} c_{n}^\alpha \phi^G_{nl}(r),
\end{equation} 
where $\alpha$ refers to the channel quantum numbers. The coefficients
$c_{n}^{\alpha}$ and the eigenenergy $E$ are determined from the Rayleigh-Ritz
variational principle
\begin{equation}
\sum_{n=1}^{n_{max}} \left[ \left(T_{n'n}^\alpha-EN_{n'n}^\alpha\right)
c_{n}^\alpha+\sum_{\alpha'} \ V_{n'n}^{\alpha\alpha'}c_{n}^{\alpha'}=0 \right],
\end{equation}
where $T_{n'n}^\alpha$, $N_{n'n}^\alpha$ and $V_{n'n}^{\alpha\alpha'}$ are the 
matrix elements of the kinetic energy, the normalization and the potential, 
respectively. $T_{n'n}^\alpha$ and $N_{n'n}^\alpha$ are diagonal whereas the
mixing between different channels is given by $V_{n'n}^{\alpha\alpha'}$.

Following Ref.~\cite{Hiyama2003} we employ Gaussian trial functions whose ranges are
in geometric progression. This is useful in optimizing the ranges with a small
number of free parameters. Moreover, this distribution of range parameters is
dense at small ranges which is well suited for making the wave function correlated
with short range potentials. The fast damping of the gaussian tail is not a real
problem since we can choose the maximal range much larger than the hadronic size.

Table~\ref{tab:parameters} shows the model parameters fitted over all spectrum 
of mesons and needed for the heavy quark sector. Note that, once the parameters
are fitted, our confinement interaction is dominantly scalar.

\begin{table}[t!]
\begin{center}
\begin{tabular}{c|cc}
\hline
\hline
Quark mass & $m_{c}$ (MeV) & $1763$ \\
\hline
Confinement & $a_{c}$ (MeV) & $507.4$ \\
	    & $\mu_{c}$ $(\mbox{fm}^{-1})$ & $0.576$ \\
	    & $\Delta$ (MeV) & $184.432$ \\
	    & $a_{s}$ & $0.81$ \\
\hline
OGE & $\alpha_{0}$ & $2.118$ \\
    & $\Lambda_{0}$ $(\mbox{fm}^{-1})$ & $0.113$ \\
    & $\mu_{0}$ (MeV) & $36.976$ \\
    & $\hat{r}_{0}$ (fm) & $0.181$ \\
    & $\hat{r}_{g}$ (fm) & $0.259$ \\
\hline
\hline
\end{tabular}
\caption{\label{tab:parameters} Quark model parameters.}
\end{center}
\end{table}

Finally, Table~\ref{tab:predmasses} shows the masses predicted by our model for
the vector charmonium states, the comparison with the experimental data and some
possible assignments of $XYZ$ mesons. Further details on the spectrum and other
properties of vector charmonium states can be found in Ref.~\cite{Segovia2008}. 

\begin{table}[t!]
\begin{center}
\begin{tabular}{ccccc}
\hline
\hline
$J^{PC}$ & $n$ & ${\rm M}_{\rm The.}$ & ${\rm M}_{\rm Exp.}$ & \\
\hline
$1^{--}$ & $1$ & $3096$ & $3096.916\pm0.011$ & \cite{PDG2012} \\
& $2$ & $3703$ & $3686.108^{+0.011}_{-0.014}$ & \cite{PDG2012} \\
& $3$ & $3796$ & $3778.1\pm1.2$ & \cite{PDG2012} \\
& $4$ & $4097$ & $4039\pm1$ & \cite{PDG2012} \\
& $5$ & $4153$ & $4153\pm3$ & \cite{PDG2012} \\
& $6$ & $4389$ & $4361\pm9\pm9$ & \cite{Wang2007} \\
& $7$ & $4426$ & $4421\pm4$ & \cite{PDG2012} \\
& $8$ & $4614$ & $4634^{+8+5}_{-7-8}$ & \cite{ref4630} \\
& $9$ & $4641$ & $4664\pm11\pm5$ & \cite{Wang2007} \\
\hline
\hline
\end{tabular}
\caption{\label{tab:predmasses} Masses, in MeV, of $1^{--}$ charmonium states.
We compare with the well established states in Ref.~\cite{PDG2012} and assign
possible $XYZ$ mesons.}
\end{center}
\end{table}
 
\section{Strong decays}
\label{sec:decaymodel}

The microscopic decay models are an attempt to describe strong interactions in terms
of quark and gluon degrees of freedom. As mentioned above, after the pioneering work
of Eichten {\it et al.}~\cite{Eichten78,Eichten06} which assume that the strong
decays are driven by the time-like component of the confining interaction, only few
works have addressed this topic in a partial way without discussing the relationship
between this decays and the nature of confinement.

We shall assume that the responsible for the strong decays is the full quark-quark
interaction of our model which includes one-gluon exchange, scalar and vector
confining interactions, allowing in this way the study of the influence of the
different pieces on the final results. 

The associated decay amplitudes of the one-gluon exchange and the confinement
interactions should be added coherently. Therefore, the current-current
interactions can be written in the generic form as~\cite{Swanson96}
\begin{equation}
H_{I}=\frac{1}{2}\int 
d^{3}\!xd^{3}\!y\,J^{a}(\vec{x})K(|\vec{x}-\vec{y}|)J^{a}(\vec{y}).
\label{Hint}
\end{equation}
The current $J^{a}$ in Eq.~(\ref{Hint}) is assumed to be a color octet. The
currents, $J$, with the color dependence $\lambda^{a}/2$ factored out and the
kernels, $K(r)$, for the interactions are
\begin{itemize}
\item Currents
\begin{equation}
\label{currents}
J(\vec{x})=\bar{\psi}(\vec{x})\,\Gamma\,\psi(\vec{x})= \begin{cases} 
\bar{\psi}(\vec{x})\,\mathcal{I}\,\psi(\vec{x}) & \mbox{Scalar Lorentz current,}
\\
\bar{\psi}(\vec{x})\,\gamma^{0}\,\psi(\vec{x}) & \mbox{Time-like vector
Lorentz current,} \\
\bar{\psi}(\vec{x})\,\vec{\gamma}\,\psi(\vec{x}) & \mbox{Space-like 
vector Lorentz current.} \end{cases}
\end{equation}
\item Kernels
\begin{equation}
\label{kernels}
K(r)=\begin{cases} 
-4a_s\left[-a_{c}(1-e^{-\mu_{c}r})+\Delta\right] 
& \mbox{Scalar Confinement,} \\ 
+4(1-a_s)\left[-a_{c}(1-e^{-\mu_{c}r})+\Delta\right] 
& \mbox{Static Vector Confinement,} \\ 
-4(1-a_s)\left[-a_{c}(1-e^{-\mu_{c}r})+\Delta\right] 
& \mbox{Transverse Vector Confinement,} \\ 
+\frac{\alpha_{s}}{r} & \mbox{Coulomb OGE,} \\
-\frac{\alpha_{s}}{r} & \mbox{Transverse OGE.} \end{cases}
\end{equation}
\end{itemize}

Following Ref.~\cite{Swanson96}, we refer to this general type of interaction as
a $JKJ$ and to the specific cases considered here as $sKs$, $j^{0}Kj^{0}$ and
$j^{T}Kj^{T}$ interactions. Details of the resulting matrix elements for different
cases are given in~\ref{appendixA}.

\begin{figure}[!t]
\begin{center}
\epsfig{figure=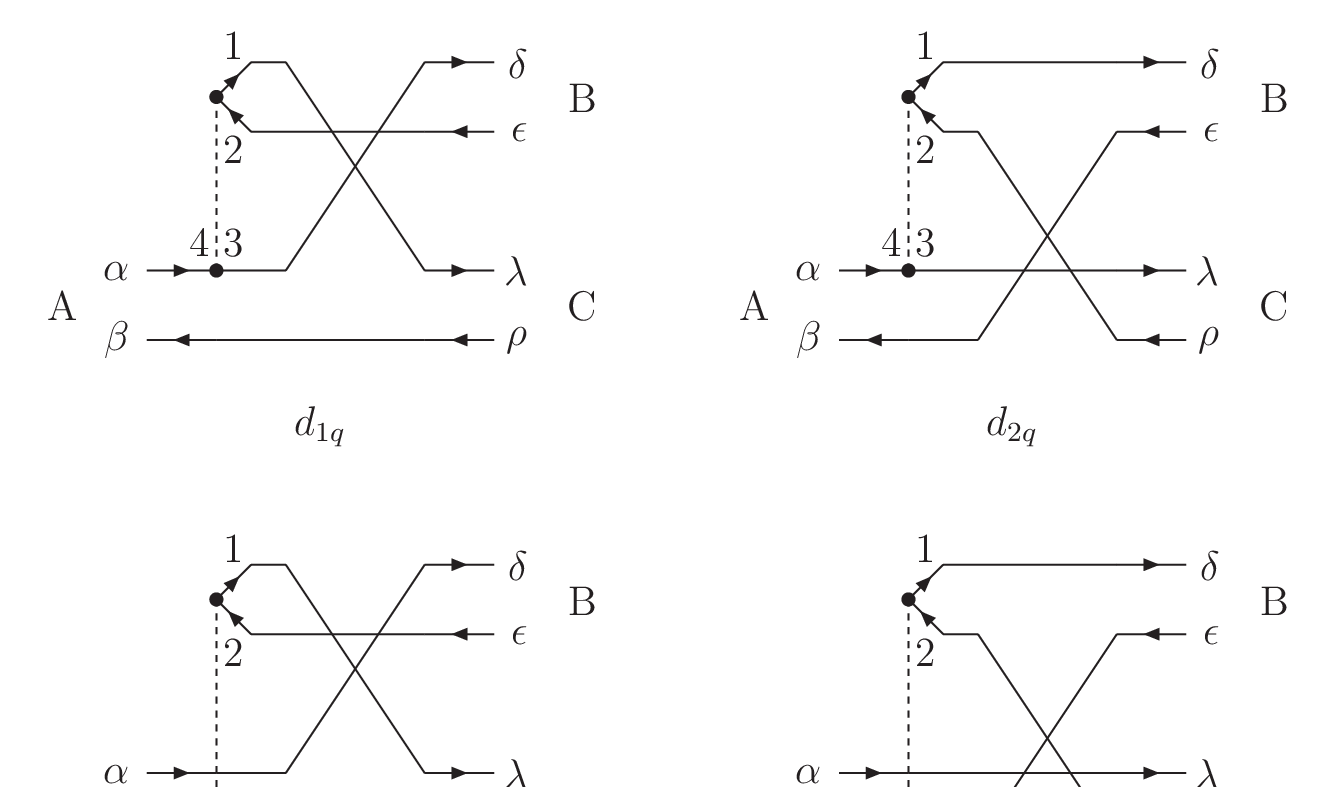,height=8.0cm,width=12.0cm}
\vspace{-0.40cm}
\caption{\label{fig:Micdiagrams} Diagrams that contribute to the decay width
through the microscopic model.}
\end{center}
\end{figure}

The diagrams that contribute to the strong decay $A\to B+C$ are shown in
Fig.~\ref{fig:Micdiagrams}. There are two coming from the quark line, $d_{1q}$
and $d_{2q}$. The difference between them is the rearrangement of the quarks and
antiquarks in the final mesons. The other two diagrams are referred to the antiquark
line, $d_{1\bar{q}}$ and $d_{2\bar{q}}$.

The total width is the sum over the partial widths characterized by the total spin,
$J_{BC}$, and the relative angular momentum, $l$, of the final mesons $B$ and $C$
\begin{equation}
\Gamma_{A\rightarrow BC}=\sum_{J_{BC},l}\Gamma_{A\rightarrow BC}(J_{BC},l),
\end{equation}
where
\begin{equation}
\Gamma_{A\rightarrow BC}(J_{BC},l)=2\pi\int 
dk_{0}\delta(E_{A}-E_{BC})|\mathcal{M}_{A\rightarrow BC}(k_{0})|^{2}
\end{equation}
and $\mathcal{M}_{A\rightarrow BC}(k_{0})$ is calculated following~\ref{appendixA}.

Using relativistic phase-space we arrive at
\begin{equation}
\begin{split}
\Gamma_{A\rightarrow 
BC}(J_{BC},l)=2\pi\frac{E_{B}E_{C}}{m_
{A}k_{0}}|\mathcal{M}_{A\rightarrow BC}(k_{0})|^{2},
\end{split}
\end{equation}
where
\begin{equation}
k_{0}=\frac{\sqrt{[m_{A}^{2}-(m_{B}-m_{C})^{2}][m_{A}^{2}-(m_{B}+m_{C})^{2}]}}{
2m_{A}}
\end{equation}
is the on-shell relative momentum of mesons $B$ and $C$.

We will compare our results with the widely use $^{3}P_{0}$ decay model which is a
particular case of Eq.~(\ref{kernels}) with only a constant scalar term
$S_{0}=\frac{3^{5/2}}{2^{4}}m_{q}\gamma$. A complete development of this model can
be found in Ref.~\cite{Bonnaz1999}.

\begin{table}[t!]
\begin{center}
\scalebox{0.90}{\begin{tabular}{cccrrrrc}
\hline
\hline
Meson & State & Channel & $\Gamma_{^{3}P_{0}}$ & ${\mathcal B}_{^3P_0}$ &
$\Gamma_{\rm Mic.}$ & ${\mathcal B}_{\rm Mic.}$ \\
\hline
$\psi(3770)$ & $1^{3}D_{1}$ & $D^{+}D^{-}$ & $11.34$ & $42.8$ & $8.03$ & $42.3$
\\
& & $D^{0}\bar{D}^{0}$ & $15.13$ & $57.2$ & $10.94$ & $57.7$ \\
& & $DD$ & $26.47$ & $100$ & $18.97$ & $100$ \\
$27.5\pm0.9$ & & total & $26.47$ & & $18.97$ & \\[2ex]
$\psi(4040)$ & $3^{3}S_{1}$ & $DD$ & $4.61$ & $4.1$ & $10.17$ & $26.0$ \\
& & $DD^{\ast}$ & $22.23$ & $20.0$ & $18.75$ & $47.9$ \\
& & $D^{\ast}D^{\ast}$ & $82.35$ & $74.0$ & $9.06$ & $23.2$ \\
& & $D_{s}D_{s}$ & $2.08$ & $1.9$ & $1.14$ & $2.9$ \\
$80\pm10$ & & total & $111.27$ & & $39.12$ & \\[2ex]
$\psi(4160)$ & $2^{3}D_{1}$ & $DD$ & $22.82$ & $19.7$ & $17.03$ & $52.1$ \\
& & $DD^{\ast}$ & $2.22$ & $1.9$ & $7.38$ & $22.6$ \\
& & $D^{\ast}D^{\ast}$ & $83.73$ & $72.2$ & $5.28$ & $16.2$ \\
& & $D_{s}D_{s}$ & $0.24$ & $0.2$ & $2.61$ & $7.9$ \\
& & $D_{s}D_{s}^{\ast}$ & $6.94$ & $6.0$ & $0.40$ & $1.2$ \\
$103\pm8$ & & total & $115.95$ & & $32.70$ & \\[2ex]
$X(4360)$ & $4^{3}S_{1}$ & $DD$ & $8.02$ & $7.0$ & $5.73$ & $5.6$ \\
& & $DD^{\ast}$ & $8.19$ & $7.2$ & $29.81$ & $29.2$ \\
& & $D^{\ast}D^{\ast}$ & $8.87$ & $7.8$ & $46.46$ & $45.5$ \\
& & $DD_{1}$ & $54.51$ & $47.8$ & $2.18$ & $2.1$ \\
& & $DD'_{1}$ & $4.29$ & $3.8$ & $12.02$ & $11.7$ \\
& & $DD_{2}^{\ast}$ & $27.17$ & $23.8$ & $0.56$ & $0.6$ \\
& & $D_{s}D_{s}$ & $0.07$ & $0.1$ & $1.86$ & $1.8$ \\
& & $D_{s}D_{s}^{\ast}$ & $1.90$ & $1.7$ & $3.36$ & $3.3$ \\
& & $D_{s}^{\ast}D_{s}^{\ast}$ & $0.91$ & $0.8$ & $0.17$ & $0.2$ \\
$74\pm15\pm10$ & & total & $113.92$ & & $102.15$ & \\[2ex]
$\psi(4415)$ & $3^{3}D_{1}$ & $DD$ & $15.11$ & $9.5$ & $7.93$ & $18.5$ \\
& & $DD^{\ast}$ & $5.82$ &$ 3.7$ & $6.66$ & $15.6$ \\
& & $D^{\ast}D^{\ast}$ & $32.56$ & $20.5$ & $7.23$ & $16.9$ \\
& & $DD_{1}$ & $64.77$ & $40.7$ & $6.06$ & $14.2$ \\
& & $DD'_{1}$ & $6.92$ & $4.4$ & $2.12$ & $5.0$ \\
& & $DD_{2}^{\ast}$ & $23.60$ & $14.8$ & $1.82$ & $4.3$ \\
& & $D^{\ast}D_{0}^{\ast}$ & $7.12$ & $4.5$ & $2.39$ & $5.6$ \\
& & $D_{s}D_{s}$ & $0.31$ & $0.2$ & $2.22$ & $5.2$ \\
& & $D_{s}D_{s}^{\ast}$ & $0.68$ & $0.4$ & $1.09$ & $2.5$ \\
& & $D_{s}^{\ast}D_{s}^{\ast}$ & $2.13$ & $1.3$ & $5.20$ & $12.2$ \\
$62\pm20$ & & total & $159.01$ & & $42.72$ & \\
\hline
\hline
\end{tabular}}
\caption{\label{tab:strong1} Open-flavor strong decay widths, in MeV, and
branchings, in $\%$, of $\psi$ states.}
\end{center}
\end{table}
\begin{table}[t!]
\begin{center}
\scalebox{0.80}{\begin{tabular}{cccrrrr}
\hline
\hline
Meson & State & Channel & $\Gamma_{^{3}P_{0}}$ & ${\mathcal B}_{^3P_0}$ & 
$\Gamma_{\rm Mic.}$ & ${\mathcal B}_{\rm Mic.}$\\
\hline
$X(4630)$ & $5^{3}S_{1}$ & $DD$ & $6.62$ & $3.2$ & $1.44$ & $0.8$ \\
& & $DD^{\ast}$ & $26.23$ & $12.7$ & $15.82$ & $8.4$ \\
& & $D^{\ast}D^{\ast}$ & $15.57$ & $7.5$ & $30.40$ & $16.2$ \\
& & $DD_{1}$ & $2.88$ & $1.4$ & $18.70$ & $9.9$ \\
& & $DD'_{1}$ & $4.52$ & $2.2$ & $2.58$ & $1.4$ \\
& & $DD_{2}^{\ast}$ & $0.00$ & $0.0$ & $21.14$ & $11.2$ \\
& & $D^{\ast}D_{0}^{\ast}$ & $6.97$ & $3.4$ & $10.10$ & $5.4$ \\
& & $D^{\ast}D_{1}$ & $39.21$ & $19.0$ & $22.47$ & $11.9$ \\
& & $D^{\ast}D'_{1}$ & $14.35$ & $7.0$ & $26.24$ & $13.9$ \\
& & $D^{\ast}D_{2}^{\ast}$ & $80.47$ & $39.0$ & $18.28$ & $9.7$ \\
& & $D_{s}D_{s}$ & $0.92$ & $0.4$ & $1.28$ & $0.7$ \\
& & $D_{s}D_{s}^{\ast}$ & $0.30$ & $0.1$ & $6.70$ & $3.6$ \\
& & $D_{s}^{\ast}D_{s}^{\ast}$ & $1.14$ & $0.6$ & $6.34$ & $3.4$ \\
& & $D_{s}D_{s1}$ & $2.82$ & $1.4$ & $0.92$ & $0.5$ \\
& & $D_{s}D'_{s1}$ & $0.79$ & $0.4$ & $0.03$ & $0.0$ \\
& & $D_{s}D_{s2}^{\ast}$ & $0.19$ & $0.1$ & $0.22$ & $0.1$ \\
& & $D_{s}^{\ast}D_{s0}^{\ast}$ & $2.76$ & $1.3$ & $1.30$ & $0.7$ \\
& & $D_{s}^{\ast}D_{s1}$ & $0.14$ & $0.1$ & $3.74$ & $2.0$ \\
& & $D_{s}^{\ast}D'_{s1}$ & $0.26$ & $0.1$ & $0.29$ & $0.1$ \\
& & $D_{s0}^{\ast}D_{s0}^{\ast}$ & $0.22$ & $0.1$ & $0.23$ & $0.1$ \\
$92^{+40+10}_{-24-21}$ & & total & $206.37$ & & $188.22$ & \\[2ex]
$X(4660)$ & $4^{3}D_{1}$ & $DD$ & $10.92$ & $8.1$ & $3.21$ & $2.3$ \\
& & $DD^{\ast}$ & $7.55$ & $5.6$ & $4.10$ & $2.9$ \\
& & $D^{\ast}D^{\ast}$ & $38.04$ & $28.2$ & $2.67$ & $1.9$ \\
& & $DD_{1}$ & $2.41$ & $1.8$ & $20.51$ & $14.4$ \\
& & $DD'_{1}$ & $0.51$ & $0.4$ & $2.62$ & $1.8$ \\
& & $DD_{2}^{\ast}$ & $0.00$ & $0.0$ & $6.75$ & $4.8$ \\
& & $D^{\ast}D_{0}^{\ast}$ & $3.44$ & $2.5$ & $0.71$ & $0.5$ \\
& & $D^{\ast}D_{1}$ & $34.83$ & $25.8$ & $10.89$ & $7.7$ \\
& & $D^{\ast}D'_{1}$ & $6.98$ & $5.1$ & $2.96$ & $2.1$ \\
& & $D^{\ast}D_{2}^{\ast}$ & $21.92$ & $16.2$ & $77.52$ & $54.5$ \\
& & $D_{s}D_{s}$ & $0.96$ & $0.7$ & $1.46$ & $1.0$ \\
& & $D_{s}D_{s}^{\ast}$ & $0.00$ & $0.0$ & $1.35$ & $0.9$ \\
& & $D_{s}^{\ast}D_{s}^{\ast}$ & $0.33$ & $0.2$ & $4.28$ & $3.0$ \\
& & $D_{s}D_{s1}$ & $3.63$ & $2.7$ & $0.0$ & $0.0$ \\
& & $D_{s}D'_{s1}$ & $1.09$ & $0.8$ & $0.62$ & $0.4$ \\
& & $D_{s}D_{s2}^{\ast}$ & $0.08$ & $0.1$ & $0.07$ & $0.1$ \\
& & $D_{s}^{\ast}D_{s0}^{\ast}$ & $1.18$ & $0.9$ & $0.43$ & $0.3$ \\
& & $D_{s}^{\ast}D_{s1}$ & $0.48$ & $0.4$ & $0.93$ & $0.6$ \\
& & $D_{s}^{\ast}D'_{s1}$ & $0.17$ & $0.1$ & $0.37$ & $0.3$ \\
& & $D_{s0}^{\ast}D_{s0}^{\ast}$ & $0.53$ & $0.4$ & $0.74$ & $0.5$ \\
$48\pm15\pm3$ & & total & $135.06$ & & $142.19$ & \\
\hline
\hline
\end{tabular}}
\caption{\label{tab:strong2} Open-flavor strong decay widths, in MeV, and
branchings, in $\%$, of $\psi$ states (Continuation).}
\end{center}
\end{table}

\section{Results}
\label{sec:comparison}

From an experimental point of view there are a few data in the open-charm decays of
the $1^{--}$ $c\bar{c}$ resonances. The main experimental data are the resonance
parameters, mass and total decay width, of the excited $\psi$ states.

Tables~\ref{tab:strong1} and~\ref{tab:strong2} show the strong decay widths predicted
by the microscopic model for the $1^{--}$ $c\bar{c}$ states established in
Table~\ref{tab:predmasses} compared with the experimental data and the $^{3}P_{0}$
results. The parameter $\gamma$ of the $^{3}P_{0}$ model has been fitted in
Ref.~\cite{Segovia2012}. The notation $D_{1}D_{2}$ includes the $D_{1}\bar{D}_{2}$
and $\bar{D}_{1}D_{2}$ combination of well defined $C\!P$ quantum numbers. For the
kinematics we use experimental masses whenever they are available.

One can see that the total decay widths predicted by the microscopic model are in
general lower than the experimental ones, whereas those predicted by the $^{3}P_{0}$
model reproduce the data in a better way. However, it is worth to notice that the
correct order of magnitude of the strong decays is given by the microscopic model
with no free parameter in contrast with the one parameter of the $^{3}P_{0}$ model.

When we go up through the spectrum, the states are more and more wide and the total
widths for $S$ and $D$-waves are larger in both decay models, always $D$-wave widths
are smaller.

In order to disentangle the contribution of the different quark-quark potential 
pieces, we compare in Table~\ref{tab:compj0Kj0} the results of the full model with
the ones taken into account only the time-like component of the confinement
potential. These last results can be compared with those of Ref.~\cite{Eichten06}
which includes the same pieces of the current although in a slightly different model.

It seems that results including only time-like vector confinement component are
better than those of the full model. However, if one looks to the ratios between the
different decay channels (Table~\ref{tab:ratios}) none of the models are able to
reproduce the experimental data. This fact suggest that the dynamics of the
charmonium strong decays is far from be a simple process and more Fock components of
the wave function can be involved in the decay~\cite{entem2011molecular}.

\begin{table}[t!]
\begin{center}
\begin{tabular}{lcccc}
\hline
\hline
Decay & Ref.~\cite{Eichten06} & $j^{0}Kj^{0}$ & Mic. & Exp.~\cite{PDG2012} \\
\hline
$\psi(3770)\to DD$ & $20.1$ & $29.8$ & $19.0$ & $25.6\pm3.4$ \\
total & $20.1$ & $29.8$ & $19.0$ & $27.5\pm0.9$ \\
\hline
$\psi(4040)\to DD$ & $0.1$ & $1.4$ & $10.2$ & \\
$\psi(4040)\to DD^{\ast}$ & $33.0$ & $25.2$ & $18.7$ & \\
$\psi(4040)\to D^{\ast}D^{\ast}$ & $33.0$ & $35.0$ & $9.1$ & \\
$\psi(4040)\to D_{s}D_{s}$ & $8.0$ & $0.3$ & $1.1$ & \\
total & $74.0$ & $61.9$ & $39.1$ & $80\pm10$ \\
\hline
$\psi(4160)\to DD$ & $3.2$ & $25.0$ & $17.0$ & \\
$\psi(4160)\to DD^{\ast}$ & $6.9$ & $0.5$ & $7.4$ & \\
$\psi(4160)\to D^{\ast}D^{\ast}$ & $41.9$ & $21.3$ & $5.3$ & \\
$\psi(4160)\to D_{s}D_{s}$ & $5.6$ & $0.03$ & $2.6$ & \\
$\psi(4160)\to D_{s}D_{s}^{\ast}$ & $11.0$ & $0.6$ & $0.4$ & \\
total & $69.2$ & $47.4$ & $32.7$ & $103\pm8$ \\
\hline
\hline
\end{tabular}
\caption{\label{tab:compj0Kj0} Open-flavor strong decay widths, in MeV, of
$\psi$ states reported in Ref.~\cite{Eichten06} and our decay rates taking into
account the static vector contribution or the full model.}
\end{center}
\end{table}

\begin{table}[t!]
\begin{center}
\begin{tabular}{cccccccc}
\hline
\hline
State & Ratio & $j^{0}Kj^{0}$ & Mic. & $^{3}P_{0}$ & Ref.~\cite{Eichten06}
& Measured~\cite{PDG2012} \\
\hline
\tstrut
$\psi(4040)$ & $DD/DD^{\ast}$ & $0.06$ & $0.54$ & $0.21$ & 0.003&$0.24\pm0.05\pm0.12$
\\
& $D^{\ast}D^{\ast}/DD^{\ast}$ & $1.39$ & $0.48$ & $3.70$ & 1.0&$0.18\pm0.14\pm0.03$
\\
$\psi(4160)$ & $DD/D^{\ast}D^{\ast}$ & $1.17$ & $3.23$ & $0.27$ & 0.076&
$0.02\pm0.03\pm0.02$ \\
& $DD^{\ast}/D^{\ast}D^{\ast}$ & $0.02$ & $1.40$ & $0.03$ & 0.16 &$0.34\pm0.14\pm0.05$
\\
$\psi(4415)$ & $DD/D^{\ast}D^{\ast}$ & $1.54$ & $1.10$ & $0.46$ & - &
$0.14\pm0.12\pm0.03$ \\
& $DD^{\ast}/D^{\ast}D^{\ast}$ & $0.28$ & $0.92$ & $0.18$ & - &$0.17\pm0.25\pm0.03$
\\
\hline
\hline 
\end{tabular}
\caption{\label{tab:ratios}  Open-flavor strong ratios of $\psi$ states predicted by
different decay models and their comparison with the experimental data.}
\end{center}
\end{table}

\section{Conclusions}
\label{sec:conclusions}

Microscopic models of meson strong decays into two mesons depend on the transition
Hamiltonian which drives the decay mechanism. We have developed a model in which the
full Hamiltonian that determines the spectrum is used for the decay. In general, the
obtained total decay widths are lower than the experimental data although the
order of magnitude is reproduced without any free parameter.

It seems that, considering only the confinement time-like components, the agreement 
with the experimental data is improved. This fact seems to be in line with the
conclusions of Ref.~\cite{PhysRevD.55.1578}. The authors stated that the Dirac
structure of confinement should be of a time-like nature which dynamically generates
an effective scalar interaction as required by the hadron spectroscopy. However, fine
details of the charmonium decays, like the ratios between different decay channels,
are not reproduced by any model which suggests that the strong decays into charmed
mesons is still an open problem. 

\section*{Acknowledgements}

This work has been partially funded by Ministerio de Ciencia y Tecnolog\'ia
under Contract No. FPA2010-21750-C02-02, by the European Community-Research
Infrastructure Integrating Activity 'Study of Strongly Interacting Matter'
(HadronPhysics3 Grant No. 283286), by the Spanish Ingenio-Consolider 2010
Program CPAN (CSD2007-00042) and also, in part, by the U.S. Department of Energy,
Office of Nuclear Physics, under contract DE-AC02-06CH11357.

\appendix

\section{Matrix elements in the microscopic model}
\label{appendixA}

\subsection{Transition operator}

If one considers only the contributions in which a quark-antiquark pair is created,
the interaction Hamiltonian in Eq.~(\ref{Hint}) reduces to the following transition
operator
\begin{equation}
\begin{split}
T &= \int d^{3}\!x d^{3}\!y \, \frac{1}{2} \, K(|\vec{x}-\vec{y}|)
\int\frac{d^{3}\!p_{1}d^{3}\!p_{2}d^{3}\!p_{3}d^{3}\!p_{4}}{(2\pi)^{6}}
\sqrt{\frac{m_{1}m_{2}m_{3}m_{4}}{E_{\vec{p}_{1}}E_{\vec{p}_{2}}E_{ \vec{p}_{3}}
E_{\vec{p}_{4}}}} \sum_ {r_ {1}}\sum_{r_{2}}\sum_{r_{3}}\sum_{r_{4}} \\
&
\left[ +b_{r_{1}}(\vec{p}_{1})b^{\dagger}_{r_{2}}(\vec{p}_{2})
a^{\dagger}_{r_{3}}(\vec{p}_{3})b^{\dagger}_{r_{4}}(\vec{p}_{4})\left[\bar{v}_{
r_{1}}(\vec{p}_{1})\Gamma
v_{r_{2}}(\vec{p}_{2})\right]\left[\bar{u}_{r_{3}}(\vec{p}_{3})\Gamma
v_{r_{4}}(\vec{p}_{4})\right]e^{+i(\vec{p}_{1}-\vec{p}_{2})\cdot\vec{x}}e^{
-i(\vec{p}_{3}+\vec{p}_{4})\cdot\vec{y}} \right. \\
&
\,\,\,+a^{\dagger}_{r_{1}}(\vec{p}_{1})a_{r_{2}}(\vec{p}_{2})
a^{\dagger}_{r_{3}}(\vec{p}_{3})b^{\dagger}_{r_{4}}(\vec{p}_{4})\left[\bar{u}_{
r_{1}}(\vec{p}_{1})\Gamma
u_{r_{2}}(\vec{p}_{2})\right]\left[\bar{u}_{r_{3}}(\vec{p}_{3})\Gamma
v_{r_{4}}(\vec{p}_{4})\right]e^{-i(\vec{p}_{1}-\vec{p}_{2})\cdot\vec{x}}e^{
-i(\vec{p}_{3}+\vec{p}_{4})\cdot\vec{y}} \\
&
\,\,\,+a^{\dagger}_{r_{1}}(\vec{p}_{1})b^{\dagger}_{r_{2}}(\vec{p}_{2})
b_{r_{3}}(\vec{p}_{3})b^{\dagger}_{r_{4}}(\vec{p}_{4})\left[\bar{u}_{r_{1}}(\vec
{p}_{1})\Gamma
v_{r_{2}}(\vec{p}_{2})\right]\left[\bar{v}_{r_{3}}(\vec{p}_{3})\Gamma
v_{r_{4}}(\vec{p}_{4})\right]e^{-i(\vec{p}_{1}+\vec{p}_{2})\cdot\vec{x}}e^{
+i(\vec{p}_{3}-\vec{p}_{4})\cdot\vec{y}} \\
&
\left.\,\,\,+a^{\dagger}_{r_{1}}(\vec{p}_{1})b^{\dagger}_{r_{2}}(\vec{p}_{2})
a^{\dagger}_{r_{3}}(\vec{p}_{3})a_{r_{4}}(\vec{p}_{4})\left[\bar{u}_{r_{1}}(\vec
{p}_{1})\Gamma
v_{r_{2}}(\vec{p}_{2})\right]\left[\bar{u}_{r_{3}}(\vec{p}_{3})\Gamma
u_{r_{4}}(\vec{p}_{4})\right]e^{-i(\vec{p}_{1}+\vec{p}_{2})\cdot\vec{x}}e^{
-i(\vec{p}_{3}-\vec{p}_{4})\cdot\vec{y}}\,\right],
\end{split}
\end{equation}
where the first term is equal to the third one. This can be seen exchanging the
$\vec{x}$ and $\vec{y}$ variables in the first term and then, changing
$1\leftrightarrow3$ and $2\leftrightarrow4$ particles taking into account the
anti-commutation rules of the creation and destruction operators to arrive to the
third term. This is possible because the kernel depends on $\vec{x}$ and $\vec{y}$ as
$|\vec{x}-\vec{y}|$. The same occurs with the second and fourth terms. Therefore we
have a factor two and we can write the transition operator as
\begin{equation}
\begin{split}
T &= \int d^{3}\!x d^{3}\!y\,K(|\vec{x}-\vec{y}|)
\int\frac{d^{3}\!p_{1}d^{3}\!p_{2}d^{3}\!p_{3}d^{3}\!p_{4}}{(2\pi)^{6}}
\sqrt{\frac{m_{1}m_{2}m_{3}m_{4}}{E_{\vec{p}_{1}}E_{\vec{p} _{2}}E_{\vec{p}_{3
}}E_{\vec{p}_{4}}}}\sum_{r_{1},r_{2},r_{3},r_{4}} \\
&
\left[+a^{\dagger}_{r_{1}}(\vec{p}_{1})b^{\dagger}_{r_{2}}(\vec{p}_{2}
)a^{\dagger}_{r_{3}}(\vec{p}_{3})a_{r_{4}}(\vec{p}_{4})\left[\bar{u}_{r_{1}}
(\vec{p}_{1})\Gamma
v_{r_{2}}(\vec{p}_{2})\right]\left[\bar{u}_{r_{3}}(\vec{p}_{3 })\Gamma
u_{r_{4}}(\vec{p}_{4})\right]e^{-i(\vec{p}_{1}+\vec{p}_{2})\cdot\vec{x}}e^{
-i(\vec{p}_{3}-\vec{p}_{4})\cdot\vec{y}}\right. \\
&
\left.\,\,+a^{\dagger}_{r_{1}}(\vec{p}_{1})b^{\dagger}_{r_{2}}(\vec{p}_{2})
b_{r_{3}}(\vec{p}_{3})b^{\dagger}_{r_{4}}(\vec{p}_{4})\left[\bar{u}_{r_{1}}(\vec
{p}_{1})\Gamma v_{r_{2}}(\vec{p}_{2})\right]\left[\bar{v}_{r_{3}}(\vec{p}_{3}
)\Gamma v_{r_{4}}(\vec{p}_{4})\right]e^{-i(\vec{p}_{1}+\vec{p}_{2})\cdot\vec{x}}
e^{+i(\vec{p}_{3}-\vec{p}_{4})\cdot\vec{y}}\,\right],
\end{split}
\end{equation}
where the first and second terms refer to the $q\bar{q}$ pair creation from the quark
line and from the antiquark line, respectively. The diagram representation of these
two terms can be seen in Fig.~\ref{fig:Micdiagrams}, diagrams $d_{1q}$ and
$d_{1\bar{q}}$. For illustration we build the result from the diagram $d_{1q}$, the
transition operator is
\begin{equation}
\begin{split}
T &= \int d^{3}\!x d^{3}\!y\,K(|\vec{x}-\vec{y}|)
\int\frac{d^{3}\!p_{1}d^{3}\!p_{2}d^{3}\!p_{3}d^{3 }\!p_{4}}{(2\pi)^{6}}
\sqrt{\frac{m_{1}m_{2}m_{3}m_{4}}{E_{\vec{p}_{1}}E_{\vec{p}_{2}}E_{\vec{p}_{3}}
E_{\vec{p}_{4}}}}\sum_{r_{1},r_{2},r_{3},r_{4}} \\
&
\left[\right. 
a^{\dagger}_{r_{1}}(\vec{p}_{1})b^{\dagger}_{r_{2}}(\vec{p}_{2})
a^{\dagger}_{r_{3}}(\vec{p}_{3})a_{r_{4}}(\vec{p}_{4})\left[\bar{u}_{r_{1}}(\vec
{p}_{1})\Gamma
v_{r_{2}}(\vec{p}_{2})\right]\left[\bar{u}_{r_{3}}(\vec{p}_{3})\Gamma
u_{r_{4}}(\vec{p}_{4})\right]e^{-i(\vec{p}_{1}+\vec{p}_{2})\cdot\vec{x}}e^{
-i(\vec{p}_{3}-\vec{p}_{4})\cdot\vec{y}}\left.\right]. \\
\end{split}
\end{equation}
The calculation of the diagram $d_{1\bar{q}}$ can be followed from that of the
diagram $d_{1q}$. If the initial meson is formed by a quark and an antiquark with
equal masses, the contribution of both diagrams to the decay rate is the same and
they contribute constructively.

Now, we can integrate in $\vec{x}$ and $\vec{y}$ 
\begin{equation}
\begin{split}
T= & \int d^{3}\!p_{1}d^{3}\!p_{2}d^{3}\!p_{3}d^{3 }\!p_{4} \, \tilde K(|\vec{Q}|) \,
\delta^{(3)}(\vec{p}_{1}+\vec{p}_{2}+\vec{p}_{3}-\vec{p}_{4})
\sqrt{\frac{m_{1}m_{2}m_{3}m_{4}}{E_{\vec{p}_{1}}E_{\vec{p}_{2}}E_{\vec{p}_{3}}
E_{\vec{p}_{4}}}}
\\
&
\sum_{r_{1},r_{2},r_{3},r_{4}}\left[ a^{\dagger}_{r_{1}}(\vec{p}_{1})b^{
\dagger}_{r_{2}}(\vec{p}_{2})a^{\dagger}_{r_{3}}(\vec{p}_{3})a_{r_{4}}(\vec{p}_{
4})\left[\bar{u}_{r_{1}}(\vec{p}_{1}) \Gamma v_{r_{2}}(\vec{p}_{2})\right]
\left[\bar{u}_{r_{3}}(\vec{p}_{3}) \Gamma u_{r_{4}}(\vec{p}_{4})\right] \right],
\\
\end{split}
\end{equation}
where $\vec{Q}=\vec{p}_{1}+\vec{p}_{2}=\vec{p}_{4}-\vec{p}_{3}$ is the momentum
transferred, $\tilde K(|\vec{Q}|)$ is the Fourier transform of the kernel $K(r)$
and the $\delta$-function implies the momentum conservation.

\subsection{Transition amplitude}

We are interested on the transition amplitude for the reaction $(\alpha\beta)_{A} \to
(\delta\epsilon)_{B} + (\lambda\rho)_{C}$. In the center-of-mass reference system of
meson $A$ one has $\vec{K}_{A}=\vec{K}_{0}=0$ and the matrix element factorizes as
follow
\begin{equation}
\left\langle BC|T|A\right\rangle = \delta^{(3)}(\vec{K}_{0})
\mathcal{M}_{A\rightarrow BC}.
\end{equation}

The initial state in second quantization is
\begin{equation}
\left.|A\right\rangle
=\int d^{3}p_{\alpha}d^{3}p_{\beta}\delta^{(3)}(\vec{K}_{A}-\vec{P}_{A})
\phi_{A}(\vec{p}_{A})a_{\alpha}^{\dagger}(\vec{p}_{\alpha})b_{\beta}^{\dagger}
(\vec{p}_{\beta})\left.|0\right\rangle,
\label{eq:Istate}
\end{equation}
where $\alpha$ $(\beta)$ are the spin, flavor and color quantum numbers of the quark
(antiquark). The wave function $\phi_{A}(\vec{p}_{A})$ denotes a meson $A$ in a color
singlet with an isospin $I_{A}$ with projection $M_{I_{A}}$, a total angular momentum
$J_{A}$ with projection $M_{A}$, $J_{A}$ is the coupling of angular momentum $L_{A}$
and spin $S_{A}$. The $\vec{p}_{\alpha}$ and $\vec{p}_{\beta}$ are the momentum of
quark and antiquark, respectively. The $\vec{P}_{A}$ and $\vec{p}_{A}$ are the total
and relative momentum of the $(\alpha\beta)$ quark-antiquark pair within the meson
$A$. The final state is more complicated than the initial one because it is a
two-meson state. It can be written as
\begin{equation}
\begin{split}
|BC\!\!\left.\right\rangle =& \frac{1}{\sqrt{1+\delta_{BC}}}\int d^{3}K_{B}
d^{3}K_{C}\sum_{m,M_{BC}}\left\langle\right.
\!\!J_{BC}M_{BC}lm|J_{T}M_{T}\!\!\left.\right\rangle\delta^{(3)}(\vec{K}-\vec{K_
{0}})\delta(k-k_{0}) \\ 
& 
\frac{Y_{lm}(\hat{k})}{k}\sum_{M_{B},M_{C},M_{I_{B}},M_{I_{C}}}\left\langle
J_{B}M_{B}J_{C}M_{C}|J_{BC}M_{BC}\right\rangle\left\langle
I_{B}M_{I_{B}}I_{C}M_{I_{C}}|I_{A}M_{I_{A}} \right\rangle \\ 
& 
\int d^{3}p_{\delta}d^{3}p_{\epsilon}d^{3}p_{\lambda}d^{3}p_{\rho}
\delta^{(3)}(\vec{K}_{B}-\vec{P}_{B})\delta^{(3)}(\vec{K}_{C}-\vec{P}_{C}) \\ 
&
\phi_{B}(\vec{p}_{B})\phi_{C}(\vec{p}_{C})a_{\delta}^{\dagger}(\vec{p}_{\delta}
)b_{\epsilon}^{\dagger}(\vec{p}_{\epsilon})a_{\lambda}^{\dagger}(\vec{p}_{
\lambda})b_{\rho}^{\dagger}(\vec{p}_{\rho})\left.|0\right\rangle,
\label{eq:Fstate}
\end{split}
\end{equation}
where we have followed the notation of meson $A$ for the mesons $B$ and $C$. We
assume that the final state of mesons $B$ and $C$ is a spherical wave with angular
momentum $l$. The relative and total momentum of mesons $B$ and $C$ are $\vec{k}_{0}$
and $\vec{K}_{0}$. The total spin $J_{BC}$ is obtained coupling the total angular
momentum of mesons $B$ and $C$, and $J_{T}$ is the coupling of $J_{BC}$ and $l$.

The diagrams that contribute to the reaction and are allowed by the transition
operator are shown in Fig.~\ref{fig:Micdiagrams}. Two of them are coming from the
quark line, $d_{1q}$ and $d_{2q}$, and take into account the different rearrangement
of the quarks and antiquarks in the final mesons. The other two diagrams are referred
to the antiquark line, $d_{1\bar{q}}$ and $d_{2\bar{q}}$. We have different cases:
\begin{itemize}
\item Case in which $\alpha=\mu=\bar{\beta}$. The two diagrams, $d_{1q}$ and
$d_{2q}$, contribute to the decay amplitude. The contribution of diagram $d_{1q}$ is
$M_{A\rightarrow BC}$ and the contribution from diagram $d_{2q}$ can be calculated
from the amplitude of the $d_{1q}$ diagram changing meson $B$ and
$C$ $(M_{A\rightarrow CB})$, so the total amplitude is given by
\begin{equation}
\mathcal{M}_{A\rightarrow BC}=M_{A\rightarrow
BC}+(-1)^{I_{B}+I_{C}-I_{A}+J_{B}+J_{C}-J_{BC}+l}M_{A\rightarrow CB}.
\label{eq:MCON1}
\end{equation}
\item Other case. Only one of the two diagrams contribute to the amplitude
\begin{equation}
\mathcal{M}_{A\rightarrow BC}=M_{A\rightarrow BC}.
\label{eq:MCON2}
\end{equation}
\end{itemize}

If the quark and antiquark in the original meson are the same, then the contribution
of diagram $d_{1q}$ $(d_{2q})$ is equal to the diagram $d_{1\bar{q}}$
$(d_{2\bar{q}})$ and both contribute constructively. In other case they have to be
calculated separately.

When the initial $A$ meson has definite $C$-parity we have to use final states with
definite $C$-parity. If ${\cal C}B=C$ the state has definite $C$-parity and the
amplitude is given by the above rules. If ${\cal C}B\neq C$ then the appropriate
$C$-parity combination has to be taken and this gives a factor $\sqrt{2}$ in the
amplitude (or the amplitude cancels for the wrong $C$-parity).

For illustration we build the result from the diagram $d_{1q}$ $(M_{A\rightarrow
BC})$. The amplitude is a product of a Fermi signature phase, a color factor, a
flavor factor and a spin-space overlap integral
\begin{equation}
M_{A\to BC}=\mathcal{I}_{\rm signature} \times \mathcal{I}_{\rm color}
\times \mathcal{I}_{\rm flavor} \times \mathcal{I}_{\rm spin-space}.
\end{equation}

\subsubsection{Fermi signature phase}

The Fermi signature can be read off from the diagram as the number of line
crossings because it arises from the ordering of the quark and antiquark
operators. In the case of $d_{1q}$ diagram we have 
\begin{equation}
I_{\rm signature}=(-1)^{3}=-1.
\end{equation}
\subsubsection{Color factor}

As all mesons are color singlet $q\bar{q}$ states the color term is given by
\begin{equation}
\mathcal{I}_{\rm color} = \frac{1}{3^{\frac{3}{2}}} \sum_{a} \mbox{Tr} \left[
\frac{\lambda^{a}}{2}\frac{\lambda^{a}}{2} \right]=\frac{4}{3^{\frac{3}{2}}}.
\end{equation}

\subsubsection{Flavor factor}

For the flavor sector we have
\begin{equation}
\mathcal{I}_{\rm flavor} = (-1)^{t_{\alpha}+t_{\beta}+I_{A}}
\sqrt{(2I_{B}+1)(2I_{C}+1)} \left\lbrace\begin{matrix}
t_{\beta} & I_{C} & t_{\mu} \\ I_{B} & t_{\alpha} & I_{A}
\end{matrix}\right\rbrace,
\label{eq:Isospinfactor}
\end{equation}
where $t_{\xi}$ is the isospin of a given quark or antiquark $\xi$. Note that
the isospin operator in the creation vertex is $u\bar{u}+d\bar{d}+s\bar{s}$.

\subsubsection{Spin-space factor}

The spin-space overlap integral for the diagram $d_{1q}$:
$1\leftrightarrow\mu$, $2\leftrightarrow\nu$, $3\leftrightarrow\delta'$ and
$4\leftrightarrow\alpha'$, reads as follow
\begin{equation}
\begin{split}
\mathcal{I}_{\rm spin-space}=&\frac{1}{\sqrt{1+\delta_{BC}}} \,
\sum_{m,M_{BC},M_{B},M_{C}} \left\langle J_{BC}M_{BC}lm|J_{T}M_{T}\right\rangle
\left\langle J_{B}M_{B}J_{C}M_{C}|J_{BC}M_{BC}\right\rangle \\
&
\int d^{3}\!K_{B} d^{3}\!K_{C} d^{3}\!p_{\delta} d^{3}\!p_{\epsilon}
d^{3}\!p_{\lambda} d^{3}\!p_{\rho} d^{3}\!p_{\mu} d^{3}\!p_{\nu}
d^{3}\!p_{\delta'} d^{3}\!p_{\alpha'} d^{3}\!p_{\alpha} d^{3}\!p_{\beta}
\sqrt{\frac{m_{\mu}m_{\nu}m_{\delta'}m_{\alpha'}}{E_{\vec{p}_{\mu}}E_{\vec{p}_{
\nu}}E_{\vec{p}_{\delta'}}E_{\vec{p}_{\alpha'}}}} \\
&
\delta^{(3)}(\vec{K}-\vec{K}_{0}) \delta(k-k_{0})
\delta^{(3)}(\vec{K}_{B}-\vec{P}_{B})\delta^{(3)}(\vec{ K}_{C}-\vec{P}_{C})
\delta^{(3)}(\vec{P}_{A}) \frac{Y_{lm}(\hat{k})}{k} \\
&
\phi_{B}(\vec{p}_{B})\phi_{C}(\vec{p}_{C})\phi_{A}(\vec{p}_{A}) \,
K(|\vec{p}_{\mu}+\vec{p}_{\nu}|)
\delta^{(3)}(\vec{p}_{\mu}+\vec{p}_{\nu}+\vec{p}_{\delta'}-\vec{p}_{
\alpha'}) \\
&
\sum_{\alpha',\delta',\mu,\nu}\delta_{\alpha'\alpha}\delta^{(3)}(\vec{p}_
{\alpha'}-\vec{p}_{\alpha})\delta_{\delta\delta'}\delta^{(3)}(\vec{p}_{\delta}
-\vec{p}_{\delta'})\delta_{\epsilon\nu}\delta^{(3)}(\vec{p}_{\epsilon}-\vec{p}_{
\nu})\delta_{\lambda\mu}\delta^{(3)}(\vec{p}_{\lambda}-\vec{p}_{\mu}) \\
&
\delta_{\rho\beta} \delta^{(3)}(\vec{p}_{\rho}-\vec{p}_{\beta})
\left[\bar{u}_{\mu}(\vec{p}_{\mu} )\Gamma v_{\nu}(\vec{p}_{\nu})\right]
\left[\bar{u}_{\delta'}(\vec{p}_{\delta'})\Gamma
u_{\alpha'}(\vec{p}_{\alpha'})\right].
\end{split}
\end{equation}
Now using some $\delta$-functions in momentum and spin of quarks (antiquarks), we can
simplify the above expression
\begin{equation}
\begin{split}
\mathcal{I}_{\rm spin-space}=&\frac{1}{\sqrt{1+\delta_{BC}}} \,
\sum_{m,M_{BC},M_{B},M_{C}} \left\langle J_{BC}M_{BC}lm|J_{T}M_{T}\right\rangle
\left\langle J_{B}M_{B}J_{C}M_{C}|J_{BC}M_{BC}\right\rangle \\
&
\int d^{3}\!K_{B} d^{3}\!K_{C} d^{3}\!p_{\delta} d^{3}\!p_{\rho} d^{3}\!p_{\mu}
d^{3}\!p_{\nu} d^{3}\!p_{\alpha} d^{3}\!p_{\beta}
\sqrt{\frac{m_{\mu}m_{\nu}m_{\delta}m_{\alpha}}{E_{\vec{p}_{\mu}}E_{\vec{p}_{\nu
}}E_{\vec{p}_{\delta}}E_{\vec{p}_{\alpha}}}} \\
&
\delta^{(3)}(\vec{K}-\vec{K}_{0}) \delta(k-k_{0})
\delta^{(3)}(\vec{K}_{B}-\vec{P}_{B})\delta^{(3)}(\vec{ K}_{C}-\vec{P}_{C})
\delta^{(3)}(\vec{P}_{A}) \frac{Y_{lm}(\hat{k})}{k} \\
&
\phi_{B}(\vec{p}_{B})\phi_{C}(\vec{p}_{C})\phi_{A}(\vec{p}_{A}) \,
K(|\vec{p}_{\mu}+\vec{p}_{\nu}|)
\delta^{(3)}(\vec{p}_{\delta}-(\vec{p}_{\alpha}-\vec{p}_{\mu}-\vec{p}_{\nu})) \\
&
\delta_{\rho\beta} \delta^{(3)}(\vec{p}_{\rho}-\vec{p}_{\beta})
\left[\bar{u}_{\mu}(\vec{p}_{\mu}) \Gamma v_{\nu}(\vec{p}_{\nu})\right]
\left[\bar{u}_{\delta}(\vec{p}_{\delta})\Gamma
u_{\alpha}(\vec{p}_{\alpha})\right].
\label{eq:ISSrelativistic}
\end{split}
\end{equation}
The nonrelativistic reduction of Eq.~(\ref{eq:ISSrelativistic}) without specifying
the $JKJ$ decay model is
\begin{equation}
\begin{split}
\mathcal{I}_{\rm spin-space}=&\frac{1}{\sqrt{1+\delta_{BC}}} \,
\sum_{m,M_{BC},M_{B},M_{C}} \left\langle J_{BC}M_{BC}lm|J_{T}M_{T}\right\rangle
\left\langle J_{B}M_{B}J_{C}M_{C}|J_{BC}M_{BC}\right\rangle \\
&
\int d^{3}\!K_{B} d^{3}\!K_{C} d^{3}\!p_{\delta} d^{3}\!p_{\rho} d^{3}\!p_{\mu}
d^{3}\!p_{\nu} d^{3}\!p_{\alpha} d^{3}\!p_{\beta} \,
\delta^{(3)}(\vec{K}-\vec{K}_{0}) \delta(k-k_{0}) \frac{Y_{lm}(\hat{k})}{k} \\
&
\delta^{(3)}(\vec{K}_{B}-\vec{P}_{B})\delta^{(3)}(\vec{ K}_{C}-\vec{P}_{C})
\delta^{(3)}(\vec{P}_{A}) \phi_{B}(\vec{p}_{B}) \phi_{C}(\vec{p}_{C})
\phi_{A}(\vec{p}_{A}) \\
&
 \,
K(|\vec{p}_{\mu}+\vec{p}_{\nu}|) \delta^{(3)}(\vec{p}_{\delta}-(\vec{p}_{\alpha}
-\vec{p}_{\mu}-\vec{p}_{\nu}))
\delta_{\rho\beta} \delta^{(3)}(\vec{p}_{\rho}-\vec{p}_{\beta}) \\
&
\lim_{v/c\to0}\left[\bar{u}_{\mu}(\vec{p}_{\mu}) \Gamma
v_{\nu}(\vec{p}_{\nu})\right]
\lim_{v/c\to0}\left[\bar{u}_{\delta}(\vec{p}_{\delta})\Gamma
u_{\alpha}(\vec{p}_{\alpha})\right].
\end{split}
\end{equation}

We require spin matrix elements which involve the nonrelativistic ${\cal O}(p/m)$
matrix elements of Dirac bilinears with $\Gamma=I,\,\gamma^{0},\,\vec{\gamma}$ and
Pauli spin matrix elements. These are
\begin{equation}
\begin{split}
\lim_{v/c\to0}[\bar{u}_{\mu}(\vec{p}_{\mu})\,I\,v_{\nu}(\vec{p}_{
\nu})]
&=
\frac{1}{2m_{\mu}}\,(\vec{p}_{\nu}-\vec{p}_{\mu}
)\cdot\left\langle\mu|\vec{\sigma}|\nu\right\rangle \\
&=
-\frac{1}{2m_{\mu}}\sqrt{2^{5}\pi}
\left[\mathcal{Y}_{1}\left(\frac{\vec{p}_{\mu}-\vec{p}_{\nu}}{2}
\right)\otimes\left(\frac{1}{2}\frac{1}{2}\right)1\right]_{0}, \\
\lim_{v/c\to0}[\bar{u}_{\mu}(\vec{p}_{\mu})\,\gamma^{0}\,v_{\nu}
(\vec{p}_{\nu})] 
&=
\frac{1}{2m_{\mu}}\,(\vec{p}_{\nu}+\vec{p}_{\mu}
)\cdot\left\langle\mu|\vec{\sigma}|\nu\right\rangle \\
&=
+\frac{1}{2m_{\mu}}\sqrt{2^{3}\pi}\left[\mathcal{Y}_{1}(\vec{p}_{\nu}+\vec{p}_{
\mu})\otimes\left(\frac{1}{ 2}\frac{1}{2}\right)1\right]_{0}, \\
\lim_{v/c\to0}[\bar{u}_{\mu}(\vec{p}_{\mu})\,\vec{\gamma}\,v_{\nu}
(\vec {p}_{\nu})]
&=
\left\langle\mu|\vec{\sigma}|\nu\right\rangle,
\end{split}
\end{equation}
and
\begin{equation}
\begin{split}
\lim_{v/c\to0}[\bar{u}_{\mu}(\vec{p}_{\mu})\,I\,u_{\nu}(\vec{p}_{
\nu})]
&=
\delta_{\mu\nu}, \\
\lim_{v/c\to0}[\bar{u}_{\mu}(\vec{p}_{\mu})\,\gamma^{0}\,u_{\nu}
(\vec{p}_{\nu})]
&=
\delta_{\mu\nu}, \\
\lim_{v/c\to0}[\bar{u}_{\mu}(\vec{p}_{\mu})\,\vec{\gamma}\,u_{\nu
}(\vec{p}_{\nu})]
&=
\frac{1}{2m_{\nu}}\left[(\vec{p}_{\nu}+\vec{p}_{\mu})\delta_{\mu\nu}
-i\left\langle\mu|\vec{\sigma}|\nu\right\rangle\times(\vec{p}_{
\mu}-\vec{p}_{\nu})\right],
\end{split}
\end{equation}
where we have used the relation
\begin{equation}
\vec{Y}_{1}\cdot\left\langle\vec{\sigma}\right\rangle=-\sqrt{3}
\left [ Y_ {1}
\otimes\left\langle\vec{\sigma}\right\rangle\right]_{0}=\sqrt{6}\left[Y_{1}
\otimes\left(\frac{1}{2}\frac{1}{2}\right)1\right]_{0}.
\end{equation}

Then, the expression for the different contributions are
\begin{itemize}

\item $sKs$ interaction
\begin{equation}
\begin{split}
\mathcal{I}_{\rm spin-space}^{\rm sKs} =&
\frac{-1}{\sqrt{1+\delta_{BC}}}\frac{1}{2m_{\nu}} \sqrt { 2^{5}\pi}\int
d^{3}\!K_{B}d^{3}\!K_{C}d^{3}\!p_{\alpha}d^{3}\!p_{\beta}d^{3}\!p_{\mu}d^{3}\!p_
{\nu}d^{3}\!p_{\delta} \\
&
\delta^{(3)}(\vec{K}-\vec{K}_{0})\delta^{(3)}(\vec{K}_{B}-\vec{P}_{B})\delta^{
(3)}(\vec{K}_{C}-\vec{P}_{C})\delta^{(3)}(\vec{P}_{A})\frac{\delta(k-k_{0})}{k}
\\
&
\delta^{(3)}\left(\vec{p}_{\delta}-(\vec{p}_{\alpha}-\vec{p}_{\mu}-\vec{p}_{\nu}
)\right)K(|\vec{p}_{\mu}+\vec{p}_{\nu}|) \\
&
\left\langle\right.\!\left[\left[\left[\phi_{B}(\vec{p}_{B})(s_{\alpha}s_{\nu}
)S_{B}\right]J_{B}\left[\phi_{C}(\vec{p}_{C})(s_{\mu}s_{\beta})S_{C}\right]J_{C}
\right]J_{BC}Y_{l}(\hat{k})\right] J_{T}| \\
&
|\left[\left[\phi_{A}(\vec{p}_{A})(s_{\alpha}s_{\beta})S_{A}\right]J_{A}\left[
\mathcal{Y}_{1}\left(\frac{\vec{p}_{\mu}-\vec{p}_{\nu}}{2}\right)(s_{\mu}s_{\nu}
)1\right]0 \right] J_{A}\!\left.\right\rangle.
\label{eq:ISSsKs}
\end{split}
\end{equation}

\item $j^{0}Kj^{0}$ interactions
\begin{equation}
\begin{split}
\mathcal{I}_{\rm spin-space}^{\rm j^{0}Kj^{0}} =&
\frac{1}{\sqrt{1+\delta_{BC}}}\frac{1}{2m_{\nu}} \sqrt {2^{3}\pi}\int
d^{3}\!K_{B}d^{3}\!K_{C}d^{3}\!p_{\alpha}d^{3}\!p_{\beta}d^{3}\!p_{\mu}d^{3}\!p_
{\nu}d^{3}\!p_{\delta} \\
&
\delta^{(3)}(\vec{K}-\vec{K}_{0})\delta^{(3)}(\vec{K}_{B}-\vec{P}_{B})\delta^{
(3)}(\vec{K}_{C}-\vec{P}_{C})\delta^{(3)}(\vec{P}_{A})\frac{\delta(k-k_{0})}{k}
\\
&
\delta^{(3)}\left(\vec{p}_{\delta}-(\vec{p}_{\alpha}-\vec{p}_{\mu}-\vec{p}_{\nu}
)\right)K(|\vec{p}_{\mu}+\vec{p}_{\nu}|) \\
&
\left\langle\right.\!\left[\left[\left[\phi_{B}(\vec{p}_{B})(s_{\alpha}s_{\nu}
)S_{B}\right]J_{B}\left[\phi_{C}(\vec{p}_{C})(s_{\mu}s_{\beta})S_{C}\right]J_{C}
\right]J_{BC}Y_{l}(\hat{k})\right] J_{T}| \\
&
|\left[\left[\phi_{A}(\vec{p}_{A})(s_{\alpha}s_{\beta})S_{A}\right]J_{A}\left[
\mathcal{Y}_{1}\left(\vec{p}_{\mu}+\vec{p}_{\nu}\right)(s_{\mu}s_{\nu})1\right]0
\right] J_{A}\!\left.\right\rangle.
\label{eq:ISSj0Kj0}
\end{split}
\end{equation}

\item $j^{T}Kj^{T}$ interactions
\begin{equation}
\begin{split}
\mathcal{I}_{\rm spin-space}^{\rm j^{T}Kj^{T}} =& \frac{1}{\sqrt{1+\delta_{BC}}}
\,
\sum_{m,M_{BC},M_{B},M_{C}} \left\langle J_{BC}M_{BC}lm|J_{T}M_{T}\right\rangle
\left\langle J_{B}M_{B}J_{C}M_{C}|J_{BC}M_{BC}\right\rangle \\
&
\int d^{3}\!K_{B} d^{3}\!K_{C} d^{3}\!p_{\delta} d^{3}\!p_{\rho} d^{3}\!p_{\mu}
d^{3}\!p_{\nu} d^{3}\!p_{\alpha} d^{3}\!p_{\beta} \,
\delta^{(3)}(\vec{K}-\vec{K}_{0}) \delta(k-k_{0}) \frac{Y_{lm}(\hat{k})}{k} \\
&
\delta^{(3)}(\vec{K}_{B}-\vec{P}_{B})\delta^{(3)}(\vec{ K}_{C}-\vec{P}_{C})
\delta^{(3)}(\vec{P}_{A}) \phi_{B}(\vec{p}_{B}) \phi_{C}(\vec{p}_{C})
\phi_{A}(\vec{p}_{A}) \\
&
 \,
K(|\vec{p}_{\mu}+\vec{p}_{\nu}|) \delta^{(3)}(\vec{p}_{\delta}-(\vec{p}_{\alpha}
-\vec{p}_{\mu}-\vec{p}_{\nu}))
\delta_{\rho\beta} \delta^{(3)}(\vec{p}_{\rho}-\vec{p}_{\beta}) \\
&
\lim_{v/c\to0}\left[\bar{u}_{\mu}(\vec{p}_{\mu}) \gamma^{i}
v_{\nu}(\vec{p}_{\nu})\right] \left( \delta_{ij}-\frac{Q_{i}Q_{j}}{\vec{Q}^{2}}
\right) \lim_{v/c\to0}\left[\bar{u}_{\delta}(\vec{p}_{\delta})\gamma^{j}
u_{\alpha}(\vec{p}_{\alpha})\right].
\label{eq:ISSjTKjT}
\end{split}
\end{equation}

\end{itemize}

The procedure followed to solve the above spin-space overlap integrals is
similar to that of Ref.~\cite{Bonnaz1999}.



\bibliographystyle{elsarticle-num}
\bibliography{elsarticle_micmodel}






\end{document}